\documentclass[dvips,12pt,a4paper]{article}
\usepackage{a4wide}
\usepackage{epsfig}
\usepackage{amsmath}
\usepackage{latexsym}

  \newlength{\absize}
\newcommand{\half}{{\textstyle\frac{1}{2}}}

\allowdisplaybreaks %!!!!

\catcode`@=11
\def\citer{\@ifnextchar [{\@tempswatrue\@citexr}{\@tempswafalse\@citexr[]}}
 
\def\@citexr[#1]#2{\if@filesw\immediate\write\@auxout{\string\citation{#2}}\fi
  \def\@citea{}\@cite{\@for\@citeb:=#2\do
    {\@citea\def\@citea{--\penalty\@m}\@ifundefined
       {b@\@citeb}{{\bf ?}\@warning
       {Citation `\@citeb' on page \thepage \space undefined}}%
\hbox{\csname b@\@citeb\endcsname}}}{#1}}
\catcode`@=12
\begin{document}
  \thispagestyle{empty}
  \pagestyle{empty}
  \renewcommand{\thefootnote}{\fnsymbol{footnote}}
\newpage\normalsize
    \pagestyle{plain}
    \setlength{\baselineskip}{4ex}\par
    \setcounter{footnote}{0}
    \renewcommand{\thefootnote}{\arabic{footnote}}
\newcommand{\preprint}[1]{%
  \begin{flushright}
    \setlength{\baselineskip}{3ex} #1
  \end{flushright}}
\renewcommand{\title}[1]{%
  \begin{center}
    \LARGE #1
  \end{center}\par}
\renewcommand{\author}[1]{%
  \vspace{2ex}
  {\Large
   \begin{center}
     \setlength{\baselineskip}{3ex} #1 \par
   \end{center}}}
\renewcommand{\thanks}[1]{\footnote{#1}}
\renewcommand{\abstract}[1]{%
  \vspace{2ex}
  \normalsize
  \begin{center}
    \centerline{\bf Abstract}\par
    \vspace{2ex}
    \parbox{\absize}{#1\setlength{\baselineskip}{2.5ex}\par}
  \end{center}}

\begin{flushright}
{\setlength{\baselineskip}{2ex}\par
{hep-ph/0108029} \\[2mm]
{August 2001}           \\
} %end tighter \baselineskip
\end{flushright}
\vspace*{4mm}
\vfill
\title{Characteristics of graviton-induced bremsstrahlung}
\vfill
\author{
Erik Dvergsnes,
Per Osland,
Nurcan \"Ozt\"urk}
%-----------------------------------
%   Address
%-----------------------------------
%\vspace{1cm}
\begin{center}
Department of Physics, University of Bergen, \\
     All\'{e}gaten 55, N-5007 Bergen, Norway
\end{center}
\vfill
\abstract
{We discuss Bremsstrahlung induced by graviton exchange in proton-proton
interactions at hadronic colliders, resulting from
$gg\to G\to \mu^+\mu^-\gamma$. Both the ADD and RS scenarios are discussed.
Due to the coupling of the graviton to two photons, the cross section 
has a new kinematical singularity for hard photons.
Thus, graviton-induced Bremsstrahlung tends to yield more hard photons
than QED-based Bremsstrahlung.
As compared with the corresponding two-body final state,
$\mu^+\mu^-$, the cross section is, for realistic cuts,
smaller by a factor $\sim 0.02$.
At the LHC, and with a string scale of a few TeV (ADD scenario),
or a graviton mass of a few TeV (RS scenario), a few events of high invariant
mass are expected per year.}
\vspace*{20mm}
\setcounter{footnote}{0}
\vfill

%\pagestyle{myheadings}
%\markboth{\Draft{Draft}}{\Draft{Draft}}

\newpage
    \setcounter{footnote}{0}
    \renewcommand{\thefootnote}{\arabic{footnote}}

%%%%%%%%%%%%%%%%%%%%%%%%%%%%%%%%%%%%%%%%%%%%%%%%%%%%%%%%%%%%%%%%%%%%%%%%
\section{Introduction}
%%%%%%%%%%%%%%%%%%%%%%%%%%%%%%%%%%%%%%%%%%%%%%%%%%%%%%%%%%%%%%%%%%%%%%%%

The idea of additional compact dimensions and strings at the TeV scale,
proposed by Antoniadis \cite{Antoniadis:1990ew} for solving
the hierarchy problem, together with the idea that Standard-Model (SM)
fields live on branes in a higher-dimensional space
\cite{Witten:1995ex}
have led to the even more radical
speculations that extra dimensions might be macroscopic, with SM fields
confined to the familiar
four-dimensional world (brane) \cite{Arkani-Hamed:1998rs,Randall:1999ee}.
The models which allow for gravity effects at the TeV scale
can be grouped into two kinds, those of factorizable geometry, where the extra
dimensions are macroscopic \cite{Arkani-Hamed:1998rs} (``ADD scenario''),
and those of non-factorizable (warped) geometry, with only one extra dimension 
separating ``our'' brane from a hidden
brane \cite{Randall:1999ee} (``RS scenario'').

In both these scenarios, the propagation of gravitons in the extra
dimensions leads to gravitons which from the four-dimensional
point of view are  massive. In the ADD scenario, these Kaluza--Klein (KK)
gravitons have masses starting at values of the order of milli-eV, and there
is practically a continuum of them, up to some cut-off $M_S$ (string scale)
of the order of TeV, whereas in the RS scenario they are
widely separated resonances at the TeV scale.
In both cases, they have a universal coupling to matter
and photons via the energy-momentum tensor.

These recent speculations have led to several studies 
\citer{Giudice:1999ck,Bijnens:2001gh} of 
various experimental signals induced by graviton production and exchange.
The new scenarios allow for the emission of massive gravitons
\cite{Giudice:1999ck,Mirabelli:1999rt,Han:1999sg}, which would lead to 
events with missing energy (or transverse momentum), as well as 
effects due to the exchange of virtual gravitons (instead of
photons or $Z^0$s) 
\cite{Giudice:1999ck,Han:1999sg,Hewett:1999sn,Davoudiasl:2001wi}.
These include the production of dileptons and diphotons
in electron-positron collisions, 
as well as gluon-gluon and quark-antiquark-induced
processes at the Tevatron and LHC.

In fact, several searches at LEP and the Tevatron have given direct
bounds on $M_S$ \cite{Landsberg:2000cw} of the order of a TeV, while
astrophysical arguments \cite{Hannestad:2001jv} can rule out 
$M_S$ up to a scale of around 80~TeV\footnote{This value applies
to the simplest ADD scenario, for $n=2$ extra dimensions.}.
Of course, a direct experimental search would be most worthwhile.
The above studies all focus on two-particle final states,
which are expected to be dominant, and therefore lead to
the most stringent bounds on the existence of massive gravitons.

Here, we shall investigate Bremsstrahlung induced by graviton exchange. 
While this cross section is further reduced by
${\cal O}(\alpha/\pi)$, so is the background.
It has some characteristic features resulting
from the exchange of a spin-2 particle and from the direct
graviton-photon coupling, that we would like to point out.
These features may be useful in discriminating any signal
against the background.

Specifically, we shall consider the process
\begin{equation}
\label{Eq:process-pp}
pp\to \mu^+\mu^-\gamma+X,
\end{equation}
which may get a contribution due to graviton exchange, and which
for energetic muons and photons should experimentally be 
a very clean signal.

We shall here focus on the gluon-gluon initiated sub-process
\begin{equation}
\label{Eq:process-gg}
gg\to G\to \mu^+\mu^-\gamma.
\end{equation}
There is also an annihilation process:
\begin{equation}
\label{Eq:process-qqbar}
q\bar q\to G\to \mu^+\mu^-\gamma,
\end{equation}
which will interfere with the Drell--Yan background,
but this is suppressed by a smaller convolution integral for moderate
values of the invariant mass.
At higher invariant masses, there is also a significant contribution
from quark-antiquark annihilation, as will be discussed in detail
elsewhere \cite{DOO-2}.

Since this final state is very distinct,
and since the Drell--Yan (or QED) background is well understood, 
the process (\ref{Eq:process-pp}) may offer some hope for observing 
a signal or improving on the exclusion bounds.

%%%%%%%%%%%%%%%%%%%%%%%%%%%%%%%%%%%%%%%%%%%%%%%%%%%%%%%%%%%%%%%%%%%%%%%%
\section{Two-body final states}
%%%%%%%%%%%%%%%%%%%%%%%%%%%%%%%%%%%%%%%%%%%%%%%%%%%%%%%%%%%%%%%%%%%%%%%%

The process of interest, Eq.~(\ref{Eq:process-pp}), is related to
the two-body final state
\begin{equation}
\label{Eq:pp-mumu-gaga}
pp\to \mu^+\mu^-+X, 
\end{equation}
which may proceed via gluon fusion and an intermediate graviton,
\begin{equation}
gg\to G\to \mu^+\mu^-.
\end{equation}

For massless muons, the cross section
for a single graviton exchange
is\footnote{This result is in agreement with \cite{Bijnens:2001gh}.}
\begin{equation}
\label{Eq:sigma-hat-mumu}
\hat\sigma^{(G)}_{gg\to\mu^+\mu^-}(\hat s)
=\frac{\kappa^4}{10240\pi}\,
\frac{\hat s^3}{(\hat s-m_G^2)^2+(m_G\Gamma_G)^2}, 
\end{equation}
with $\hat s=(k_1+k_2)^2$ the two-gluon invariant mass squared.
Furthermore, $m_G$ and $\Gamma_G$ are the mass and width of the graviton,
and $\kappa$ is the graviton coupling, to be defined below.
The angular distribution is given by $1-\cos^4\theta$, where $\theta$
is the c.m.\ scattering angle.

With $\xi_1$ and $\xi_2$ the fractional momenta of the two gluons,
$k_1=\xi_1 P_1$, $k_2=\xi_2 P_2$,
and $P_1$ and $P_2$ the proton momenta,
$(P_1+P_2)^2=s$,
we have
$\hat s\simeq \xi_1\xi_2s$.

For the over-all process (\ref{Eq:pp-mumu-gaga}) we thus find
\begin{equation}
\label{Eq:sigma-pp}
\sigma^{(G)}_{gg\to \mu^+\mu^-}
=\int_0^1d\xi_1\int_0^1d\xi_2\, f_g(\xi_1)f_g(\xi_2)\,
\delta\left(\xi_1\xi_2-\frac{\hat s}{s}\right)\,
\hat\sigma^{(G)}_{gg\to \mu^+\mu^-}(\hat s)
=I_{gg}(\hat s)\, 
\hat\sigma^{(G)}_{gg\to \mu^+\mu^-}(\hat s),
\end{equation}
with
\begin{equation}
\label{Eq:I_gg-convolution}
I_{gg}(\hat s)=\int_{-Y}^{Y} dy\, 
f_g\left(\sqrt{\frac{\hat s}{s}}\,e^y\right)
f_g\left(\sqrt{\frac{\hat s}{s}}\,e^{-y}\right),
\qquad Y=\half\log\frac{s}{\hat s},
\end{equation}
the relevant convolution integral over the gluon distribution functions.
For the ADD scenario, it will be more convenient to compare with
the differential cross section:
\begin{eqnarray}
\label{Eq:dsigma-pp}
\frac{d\sigma^{(G)}_{gg\to \mu^+\mu^-}}{d\hat s}
&=&\int_0^1d\xi_1\int_0^1d\xi_2\, f_g(\xi_1)f_g(\xi_2)\,
\frac{d\hat\sigma^{(G)}_{gg\to \mu^+\mu^-}}{d\hat s} \nonumber\\
&=&\int_0^1d\xi_1\int_0^1d\xi_2\, f_g(\xi_1)f_g(\xi_2)\,
\delta\left(\xi_1\xi_2s-\hat s\right)\,
\hat\sigma^{(G)}_{gg\to \mu^+\mu^-}(\hat s) \nonumber\\
&=&\frac{1}{s}I_{gg}(\hat s)\, \hat\sigma^{(G)}_{gg\to \mu^+\mu^-}(\hat s).
\end{eqnarray}

%%%%%%%%%%%%%%%%%%%%%%%%%%%%%%%%%%%%%%%%%%%%%%%%%%%%%%%%%%%%%%%%%%%%%%%%
\section{Graviton-induced Bremsstrahlung}
%%%%%%%%%%%%%%%%%%%%%%%%%%%%%%%%%%%%%%%%%%%%%%%%%%%%%%%%%%%%%%%%%%%%%%%%

The Bremsstrahlung process (\ref{Eq:process-gg})
can proceed via the four Feynman diagrams of Fig.~1,
the basic couplings for which are given by Han et al.~\cite{Han:1999sg}
(see also Giudice et al.~\cite{Giudice:1999ck}).
%%%%%%%%%%%%%%%%%%%%%%%%%%%%%%%%%%%%%%%%%%%%%%%%%%%%%%%%%%%%%%%%%%%%%%
\begin{figure}[htb]
\refstepcounter{figure}
\label{Fig:Feynman}
\addtocounter{figure}{-1}
\begin{center}
\setlength{\unitlength}{1cm}
\begin{picture}(12,7.5)
\put(0.5,0.2)
{\mbox{\epsfysize=7.7cm\epsffile{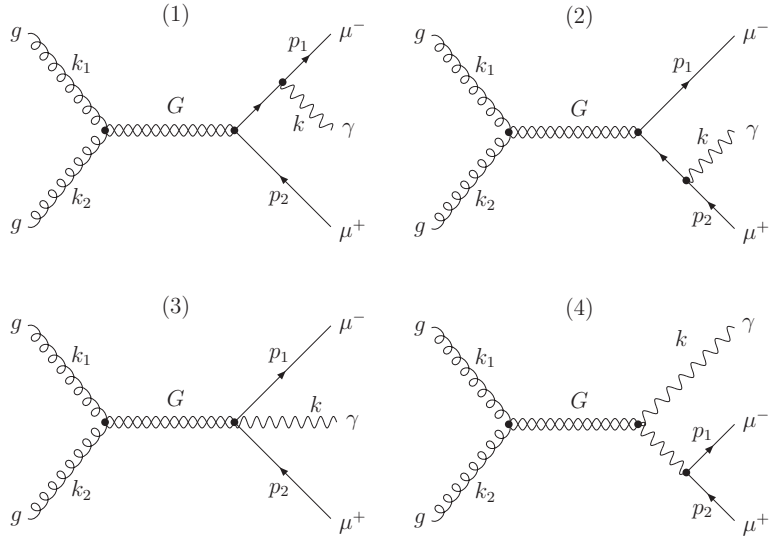}}}
\end{picture}
%\vspace*{-8mm}
\caption{Feynman diagrams for $gg\to G\to \mu^+\mu^-\gamma$.}
\end{center}
\end{figure}
%%%%%%%%%%%%%%%%%%%%%%%%%%%%%%%%%%%%%%%%%%%%%%%%%%%%%%%%%%%%%%%%%%%%%%

The evaluation of the cross section is straightforward.
We choose the (unitary) gauge ($\xi^{-1}=0$ in the notation
of \cite{Han:1999sg}), whereby the scalar field
decouples. The resulting cross section has in the
two-gluon c.m.\ frame the form (again for a single graviton exchange):
\begin{equation}
\label{Eq:dsig-3-domega}
\frac{4\pi}{d(\cos\theta)d\chi}\,
\frac{d^4\hat\sigma^{(G)}_{gg\to\mu^+\mu^-\gamma}}{dx_1dx_2}
=\frac{\alpha \kappa^4}{2560\pi^2}\,
\frac{\hat s^3}{(\hat s-m_G^2)^2+(m_G\Gamma_G)^2}\,
\frac{Z(x_1,x_2,\cos\theta,\chi)}{(1-2x_1)(1-2x_2)(1-2x_3)},
\end{equation}
where $\alpha$ is the fine-structure constant and
$x_1$, $x_2$ and $x_3$ denote the fractional energies of the muons
and the photon in the c.m.\ frame,
\begin{equation}
x_1=E_1/\sqrt{\hat s}, \qquad x_2=E_2/\sqrt{\hat s}, \qquad
x_3=\omega/\sqrt{\hat s}, \qquad 0\le x_i\le\half,
\end{equation}
with $x_1+x_2+x_3=1$,
whereas $\theta$ and $\chi$ give the orientation of the event
w.r.t.\ the gluon momenta, and
\begin{equation}
\hat s\equiv (k_1+k_2)^2=(p_1+p_2+k)^2.
\end{equation}
The denominator in (\ref{Eq:dsig-3-domega}) exhibits
the familiar singularities in the infrared and collinear limits,
$s_1\equiv(p_1+k)^2=\hat s(1-2x_2)\to0$, 
$s_2\equiv(p_2+k)^2=\hat s(1-2x_1)\to0$, as well as a
collinear singularity at
$s_3\equiv(p_1+p_2)^2=\hat s(1-2x_3)\to0$ due to the fourth Feynman diagram.

Since the underlying mechanism is the exchange of a spin-two object,
the quantity $Z$ is of fourth order in the invariants.
It is straight-forward to verify that it is gauge invariant with
respect to the gluons, as well as to the photon (actually, diagram 4
is by itself gauge invariant).
But the expression is quite lengthy, so we shall not write it out here.

The angular distribution of the non-radiative cross section 
(\ref{Eq:sigma-hat-mumu}) is given by a fourth-order polynomial in 
$\cos\theta$.
Here, just like in gluon Bremsstrahlung (see, e.g., \cite{Olsen:1980cw}),
there is an accompanying dependence on the azimuthal angle $\chi$, 
but now up to fourth order in $\cos\chi$, or, equivalently, 
up to $\cos4\chi$.

After averaging and summing over gluon, muon and photon polarizations,
and integrating over event orientations w.r.t.\ the gluon momentum, 
we find
\begin{equation}
\label{Eq:dsigma-gg}
\frac{d^2\hat\sigma^{(G)}_{gg\to\mu^+\mu^-\gamma}}{dx_1dx_2}
=\frac{\alpha \kappa^4}{2560\pi^2}\,
\frac{\hat s^3}{(\hat s-m_G^2)^2+(m_G\Gamma_G)^2}\,
\frac{Z(x_1,x_2)}{(1-2x_1)(1-2x_2)(1-2x_3)},
\end{equation}
with 
\begin{eqnarray}
\label{Eq:Z}
Z(x_1,x_2)
&=&2(x_1^2+x_2^2)[16x_1x_2-6(x_1+x_2)+3] \nonumber \\
&=&2(x_1^2+x_2^2)[1-2x_3+4x_3^2-4(x_1-x_2)^2].
\end{eqnarray}
describing the event-shape distribution.

For the over-all process (\ref{Eq:process-pp}) we thus find from 
Eq.~(\ref{Eq:sigma-pp})
\begin{equation}
\label{Eq:d2sigma-pp}
\frac{d^2\sigma^{(G)}_{gg\to\mu^+\mu^-\gamma}}{dx_1dx_2}
=I_{gg}(\hat s)\, 
\frac{d^2\hat\sigma^{(G)}_{gg\to\mu^+\mu^-\gamma}}{dx_1dx_2},
\end{equation}
with the relevant convolution integral over the gluon distribution 
functions given by Eq.~(\ref{Eq:I_gg-convolution}).
Using Eq.~(\ref{Eq:dsigma-gg}), we get
\begin{equation}
\label{Eq:d2sigma-vs-sigma-mumu}
\frac{1}{\sigma^{(G)}_{gg\to\mu^+\mu^-}}\,
\frac{d^2\sigma^{(G)}_{gg\to\mu^+\mu^-\gamma}}{dx_1dx_2}
=\frac{4\alpha}{\pi}\,
\frac{Z(x_1,x_2)}{(1-2x_1)(1-2x_2)(1-2x_3)},
\end{equation}
with $\sigma^{(G)}_{gg\to\mu^+\mu^-}$ given by 
Eqs.~(\ref{Eq:sigma-hat-mumu}) and~(\ref{Eq:sigma-pp}).

%%%%%%%%%%%%%%%%%%%%%%%%%%%%%%%%%%%%%%%%%%%%%%%%%%%%%%%%%%%%%%%%%%%%%%%%
\section{Bremsstrahlung in the ADD scenario}
%%%%%%%%%%%%%%%%%%%%%%%%%%%%%%%%%%%%%%%%%%%%%%%%%%%%%%%%%%%%%%%%%%%%%%%%

In the ADD scenario \cite{Arkani-Hamed:1998rs}, 
the coupling of each KK mode to matter is Planck-scale suppressed.
However, since the states are very closely spaced, with \cite{Han:1999sg}
\begin{equation} 
m_G^2=\frac{4\pi^2\vec n_{\rm KK}^2}{R^2},
\end{equation}
and $R/2\pi$ the compactification radii,
the coherent summation over the many modes leads to effective couplings
with strength $1/M_S$.

Explicitly, in this scenario, the graviton coupling
is in the $(4+n)$-dimensional theory given by \cite{Han:1999sg}
\begin{equation}
\hat g_{MN}=\hat \eta_{MN}+\hat \kappa \hat h_{MN}, \qquad 
\hat \kappa^2=16\pi G_{\rm N}^{(4+n)},
\end{equation}
where $G_{\rm N}^{(4+n)}$ is Newton's constant in $4+n$ dimensions.
In $4$ dimensions the coupling can be  written as 
\begin{equation}
\kappa^2=V_n^{-1} \hat \kappa^2=16\pi V_n^{-1}\,G_{\rm N}^{(4+n)}
=16\pi G_{\rm N},
\end{equation}
with $V_n$ the volume of the $n$-dimensional compactified space
($V_n=R^n$ for a torus $T^n$) and $G_{\rm N}$ the 4-dimensional 
Newton constant.

Summing coherently over all KK modes in a tower, the graviton propagator 
gets replaced \cite{Han:1999sg}:
\begin{equation}
\frac{i}{\hat s-m_G^2} \to D(\hat s)
=\frac{8\pi}{\kappa^2}\, C_4,
\end{equation}
with 
\begin{equation}
C_4\simeq
\begin{cases}
-iM_S^{-4}\log(M_S^2/\hat s), & n=2, \\
-2iM_S^{-4}/(n-2),     & n>2,
\end{cases}
\end{equation}
for $n$ extra dimensions,
and $M_S\gg\sqrt{\hat s}$ the string scale which is introduced as 
an UV cut-off.

The Bremsstrahlung cross section can thus be expressed as
\begin{equation}
\label{Eq:dsigma-gg-add}
\frac{d^3\sigma^{(G)}_{gg\to\mu^+\mu^-\gamma}}{d\hat s\,dx_1dx_2}
=\frac{\alpha \kappa^4}{2560\pi^2}\,
\frac{\hat s^3}{s}\,\left|D(\hat s)\right|^2 \,
\frac{Z(x_1,x_2)}{(1-2x_1)(1-2x_2)(1-2x_3)} \,
I_{gg}(\hat s).
\end{equation}
For a given value of $\hat s$ (characterizing the final state)
the cross section depends on three model parameters: $n$, $M_S$
and $R$, which are related as follows \cite{Han:1999sg}:
\begin{equation}
\kappa^2 R^n =8\pi(4\pi)^{n/2}\Gamma(n/2)M_S^{-(n+2)}.
\end{equation}
Relative to the two-body  cross section, the Bremsstrahlung
cross section is however {\it independent of the ADD model parameters}:
\begin{equation}
\label{Eq:add-ratio}
\left[\frac{d\sigma^{(G)}_{gg\to\mu^+\mu^-}}{d\hat s}\right]^{-1}
\frac{d^3\sigma^{(G)}_{gg\to\mu^+\mu^-\gamma}}{d\hat s\,dx_1dx_2}
=\frac{4\alpha}{\pi}\,
\frac{Z(x_1,x_2)}{(1-2x_1)(1-2x_2)(1-2x_3)},
\end{equation}
where $\sigma^{(G)}_{gg\to\mu^+\mu^-}$ refers to 
the gluon-gluon-induced two-muon cross section\footnote{This result is 
consistent with that of \cite{Hewett:1999sn}.}
\begin{equation}
\frac{d\sigma^{(G)}_{gg\to\mu^+\mu^-}}{d\hat s}
=\frac{\pi}{160}\,\frac{\hat s^3}{s}|C_4|^2\,I_{gg}(\hat s).
\end{equation}

%%%%%%%%%%%%%%%%%%%%%%%%%%%%%%%%%%%%%%%%%%%%%%%%%%%%%%%%%%%%%%%%%%%%%%%%
\section{Bremsstrahlung in the Randall--Sundrum model}
%%%%%%%%%%%%%%%%%%%%%%%%%%%%%%%%%%%%%%%%%%%%%%%%%%%%%%%%%%%%%%%%%%%%%%%%

In the Randall--Sundrum model \cite{Randall:1999ee}, 
the graviton masses are given by \cite{Davoudiasl:2000jd}
\begin{equation}
\left(m_G\right)_n=k x_n\, e^{-kr_c\pi},
\end{equation}
where $x_n$ are roots of the Bessel function of order 1,
$k$ is of the order of the (four-dimensional) Planck scale and $r_c$ 
the compactification radius of the extra 
dimension\footnote{To solve the hierarchy problem, $kr_c \sim 12$ is 
required.}.

The gravitational coupling is for the lightest KK-graviton given 
by\footnote{For simplicity we will only consider the first resonance, and let
$m_G$ refer to its mass.}
\cite{Davoudiasl:2000jd,Bijnens:2001gh}
\begin{equation}
\kappa m_G=\sqrt{2}\,x_1\, \frac{k}{\overline M_{\rm Pl}}, \qquad
\overline M_{\rm Pl}=\frac{M_{\rm Pl}}{\sqrt{8\pi}}
\simeq 2.4\times10^{18}~\text{GeV},
\end{equation}
where $x_1=3.83171$ is the first zero of the Bessel function,
$J_1(x_1)=0$.

In the Randall--Sundrum model, the gravitons are thus rather massive
and widely separated in mass.
For small values of $k/\overline M_{\rm Pl}$,
it is an excellent approximation to integrate (\ref{Eq:sigma-hat-mumu}) and 
(\ref{Eq:d2sigma-pp})
over $\hat s$
in the narrow-width approximation,
\begin{equation}
\int d\hat s
\frac{1}{(\hat s-m_G^2)^2+(m_G\Gamma_G)^2}\, f(\hat s)
\simeq \frac{\pi}{m_G\Gamma_G}\, f(m_G^2),
\end{equation}
where \cite{Han:1999sg,Allanach:2000nr}
\begin{equation}
\Gamma_G\equiv\frac{\gamma_G}{20\pi}\, m_G^3\kappa^2,
\end{equation}
with
\begin{equation}
\gamma_G=1+\chi_\gamma+\chi_Z+\chi_W+\chi_\ell+\chi_q+\chi_H, 
\end{equation}
the total graviton width in units of the two-gluon width.
Neglecting mass effects, we have \cite{Han:1999sg,Allanach:2000nr}
\begin{equation}
\chi_\gamma=\frac{1}{8}, \qquad
\chi_Z=\frac{13}{96}, \qquad
\chi_W=\frac{13}{48}, \qquad
\chi_\ell=\frac{N_\ell}{16}, \qquad
\chi_q=\frac{N_c N_q}{16}, \qquad
\chi_H=\frac{1}{48}.
\end{equation}
Here, $N_\ell=6$ is the number of leptons, and
$N_c N_q=18$ is the number of quarks weighted with color factors.

In the narrow-width approximation,
the integrated cross section (\ref{Eq:d2sigma-pp}) becomes
\begin{equation}
\label{Eq:sig-rs-int}
\int d\hat s\, \frac{d^2\sigma^{(G)}_{gg\to\mu^+\mu^-\gamma}}{dx_1dx_2}
=\frac{\alpha}{128}\, \frac{(\kappa m_G)^2}{\gamma_G}\,
I_{gg}(m_G^2)\, 
\frac{Z(x_1,x_2)}{(1-2x_1)(1-2x_2)(1-2x_3)}.
\end{equation}

%%%%%%%%%%%%%%%%%%%%%%%%%%%%%%%%%%%%%%%%%%%%%%%%%%%%%%%%%%%%%%%%%%%%%%%%
\section{Drell--Yan (QED) background}
%%%%%%%%%%%%%%%%%%%%%%%%%%%%%%%%%%%%%%%%%%%%%%%%%%%%%%%%%%%%%%%%%%%%%%%%

The same final state can also be produced by
\begin{equation}
q\bar q\to \gamma,Z^0\to \mu^+\mu^-\gamma.
\end{equation}
Photon exchange dominates over $Z^0$ exchange, and gives an elementary 
$q\bar q\to \gamma \to \mu^+\mu^-\gamma$ cross section\footnote{We 
neglect Bremsstrahlung from initial quarks, since this contribution
can be substantially reduced by suitable cuts.}
\begin{equation}
\frac{d^2\hat\sigma^{(\gamma)}_{q\bar q\to\mu^+\mu^-\gamma}}
{dx_1dx_2}
=\frac{32}{9}\,\frac{\alpha^3}{\hat s}\,
\frac{x_1^2+x_2^2}{(1-2x_1)(1-2x_2)}
\end{equation}
(divided by the quark charge squared, $Q_q^2$).
For proton-proton collisions, we get
\begin{equation}
\label{Eq:sigma-DY-pp}
\frac{d^2\sigma^{(\gamma)}_{q\bar q\to\mu^+\mu^-\gamma}}
{dx_1dx_2}
=I_{q\bar q}(\hat s)\, 
\frac{d^2\hat\sigma^{(\gamma)}_{q\bar q\to\mu^+\mu^-\gamma}}
{dx_1dx_2},
\end{equation}
with
\begin{equation}
\label{Eq:I_qq-convolution}
I_{q\bar q}(\hat s)=2\sum_q Q_q^2
\int_{-Y}^Y dy\, f_q\left(\sqrt{\frac{\hat s}{s}}\,e^y\right)
          f_{\bar q}\left(\sqrt{\frac{\hat s}{s}}\,e^{-y}\right)
\end{equation}
the appropriate quark-antiquark convolution integral for the LHC
(the factor of two accounts for the fact that either beam
can provide the quark or the antiquark).
In the ADD case we will use Eq.~(\ref{Eq:dsigma-pp}), but now with the
convolution integral from Eq.~(\ref{Eq:I_qq-convolution}), to express the
differential cross section for the QED background as
\begin{equation}
\label{Eq:dsigma-DY-pp}
\frac{d^3\sigma^{(\gamma)}_{q\bar q\to\mu^+\mu^-\gamma}}
{d\hat s\,dx_1dx_2}
=\frac{1}{s}\, I_{q\bar q}(\hat s)\, 
\frac{d^2\hat\sigma^{(\gamma)}_{q\bar q\to\mu^+\mu^-\gamma}}
{dx_1dx_2}.
\end{equation}

In the RS scenario, the signal to background ratio will depend critically on
the experimental resolution in $\hat s$, which we shall express in terms of
\begin{equation}
\delta_{\hat s}\equiv\frac{\Delta \hat s}{\hat s}
=\frac{\Delta(p_1+p_2+k)^2}{\hat s}.
\end{equation}

It is also useful to define dimensionless cross sections,
integrated over event shapes, subject to $y$-cuts:
\begin{eqnarray}
\label{Eq:sig-G-ints}
\tilde\sigma^{(G)}_{gg\to\mu^+\mu^-\gamma}
&=&\iint\limits_{s_i>y\hat s} dx_1dx_2\,
\frac{Z(x_1,x_2)}{(1-2x_1)(1-2x_2)(1-2x_3)}, \\
\label{Eq:sig-ga-ints}
\tilde\sigma^{(\gamma)}_{q\bar q\to\mu^+\mu^-\gamma}
&=&\iint\limits_{s_i>y\hat s} dx_1dx_2\,
\frac{2(x_1^2+x_2^2)}{(1-2x_1)(1-2x_2)}.
\end{eqnarray}
The $y$-cuts will remove IR soft and collinear events where the photon
has little energy, or its direction is close to that of a muon.
A milder $y$-cut, $y_3$, will remove events where the two muons
are close.
These integrated cross sections will play a crucial role when we compare
signal and background.

%%%%%%%%%%%%%%%%%%%%%%%%%%%%%%%%%%%%%%%%%%%%%%%%%%%%%%%%%%%%%%%%%%%%%%%%
\section{Event characteristics}
%%%%%%%%%%%%%%%%%%%%%%%%%%%%%%%%%%%%%%%%%%%%%%%%%%%%%%%%%%%%%%%%%%%%%%%%

It is interesting to compare the energy distribution 
for spin-two graviton exchange with the corresponding 
one resulting from the exchange of a spin-one object, like the photon
or a $Z^0$.
In Fig.~\ref{Fig:x1x2-dist} we show 
$Z(x_1,x_2)/[2(1-2x_3)(x_1^2+x_2^2)]$, which is the ratio of the integrands in
Eqs.~(\ref{Eq:sig-G-ints}) and~(\ref{Eq:sig-ga-ints}).
(The normalization is chosen such that the ratio is 1 in the IR limit,
$x_3\to0$, or $x_1\simeq x_2\to\half$.)
%%%%%%%%%%%%%%%%%%%%%%%%%%%%%%%%%%%%%%%%%%%%%%%%%%%%%%%%%%%%%%%%%%%%%%
\begin{figure}[htb]
\refstepcounter{figure}
\label{Fig:x1x2-dist}
\addtocounter{figure}{-1}
\begin{center}
\setlength{\unitlength}{1cm}
\begin{picture}(7.5,7.0)
\put(0.0,0.0)
{\mbox{\epsfysize=7.0cm\epsffile{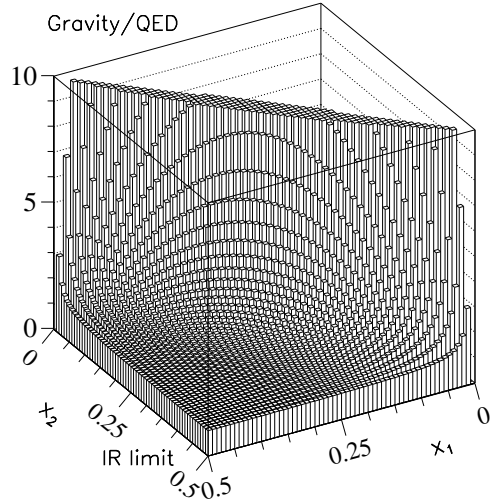}}}
\end{picture}
%\vspace*{-8mm}
\caption{Ratio of graviton-induced to QED cross section,
{\it vs.} fractional muon energies $x_1$ and $x_2$.
The ratio is normalized to 1 in the IR limit
of vanishing photon energy.}
\end{center}
\end{figure}
%%%%%%%%%%%%%%%%%%%%%%%%%%%%%%%%%%%%%%%%%%%%%%%%%%%%%%%%%%%%%%%%%%%%%%

The ``wall'' at $x_3\to\half$ is due to the collinear singularity
that arises from the fourth diagram (see Fig.~1).
Of course, when finite-mass effects are taken into account, 
the cross section is finite in this limit \cite{DOO-2}.
Integrating over some slice in $x_3$ along this ``wall'', the cross section
gets enhanced by a factor $\log(\hat s/4m_f^2)$, 
with $m_f$ the muon mass.
This partly compensates for the factor $\alpha/\pi$.

Concerning the angular distributions, we note that the photon
originating from the two-gluon initial state leads to no forward-backward
asymmetry, as is the case for the two-body final states 
\cite{Giudice:1999ck,Hewett:1999sn}.
On the other hand, with quark-antiquark initial states,
there is a forward-backward asymmetry in the photon
angular distribution in analogy with the two-body final states
\cite{Hewett:1999sn}. It could be of interest to search for this 
at the Tevatron\footnote{At the LHC, such asymmetries would be absent,
since the two beams are identical.}.

While the difference between the two cases is rather striking, 
as displayed by the ratio in Fig.~\ref{Fig:x1x2-dist},
it should be kept in mind that there are singularities along the edges
$x_1=0.5$ and $x_2=0.5$. Thus, when an integrated cross section
is considered, these singular parts (which are absent in the ratio
shown in Fig.~\ref{Fig:x1x2-dist}) play an important role,
and tend to reduce the difference between the graviton- and 
photon-exchange cases.

%%%%%%%%%%%%%%%%%%%%%%%%%%%%%%%%%%%%%%%%%%%%%%%%%%%%%%%%%%%%%%%%%%%%%%%%
\section{Discussion}
%%%%%%%%%%%%%%%%%%%%%%%%%%%%%%%%%%%%%%%%%%%%%%%%%%%%%%%%%%%%%%%%%%%%%%%%

We show in Fig.~\ref{Fig:x3cut-dist} the integrated, dimensionless
cross sections of Eqs.~(\ref{Eq:sig-G-ints}) and (\ref{Eq:sig-ga-ints}) 
vs.\ $x_3^{\rm min}$,
where we have integrated over $x_3^{\rm min}\le x_3 \le 0.5$,
subject to $y$-cuts: $s_1, s_2 \ge y \hat s$, $s_3\ge y_3 \hat s$.
Three values of the $y$-cut are considered, $y=0.01, 0.02, 0.05$,
whereas $y_3$, which controls the minimum invariant mass of the two muons, 
has been held fixed at 0.01.
At a scale $\sqrt{\hat s}=1$~TeV, the most loose cut of $y=0.01$
corresponds to muon (or photon) energies exceeding 10~GeV.
The corresponding angular cuts are well within the resolutions
foreseen at the LHC \cite{lhc}.
%%%%%%%%%%%%%%%%%%%%%%%%%%%%%%%%%%%%%%%%%%%%%%%%%%%%%%%%%%%%%%%%%%%%%%
\begin{figure}[htb]
\refstepcounter{figure}
\label{Fig:x3cut-dist}
\addtocounter{figure}{-1}
\begin{center}
\setlength{\unitlength}{1cm}
\begin{picture}(7.5,6.5)
\put(-1.0,0.0)
{\mbox{\epsfysize=7.0cm\epsffile{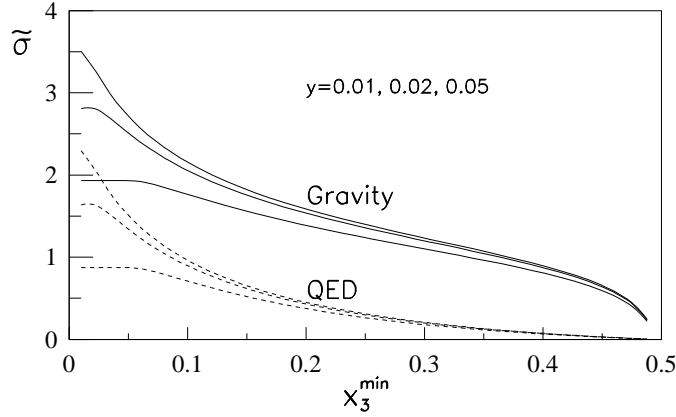}}}
\end{picture}
%\vspace*{-8mm}
\caption{Integrated dimensionless cross sections of 
Eqs.~(\ref{Eq:sig-G-ints}) and (\ref{Eq:sig-ga-ints})
vs.\ $x_3^{\rm min}$.
Solid: graviton exchange, dashed: QED.
Three values of $y$-cut are considered, as indicated,
from top to bottom.}
\end{center}
\end{figure}
%%%%%%%%%%%%%%%%%%%%%%%%%%%%%%%%%%%%%%%%%%%%%%%%%%%%%%%%%%%%%%%%%%%%%%

With the integrated dimensionless cross sections given in 
Fig.~\ref{Fig:x3cut-dist},
and the convolution integrals given by Eqs.~(\ref{Eq:I_gg-convolution})
and (\ref{Eq:I_qq-convolution}),
one may define a ``figure of merit'' as the ratio of the graviton cross
section to the Drell--Yan background.
For the ADD scenario, this will take the form
\begin{equation}
\label{Eq:R-ADD}
R^{\rm ADD}
=\frac{9}{640}\, \frac{|C_4|^2\, \hat s^4}{\alpha^2}\,
\frac{I_{gg}(\hat s)}{I_{q\bar q}(\hat s)}\,
\frac{\tilde\sigma^{(G)}_{gg\to\mu^+\mu^-\gamma}}
{\tilde\sigma^{(\gamma)}_{q\bar q\to\mu^+\mu^-\gamma}},
\end{equation}
whereas for the RS scenario (in the narrow-width approximation), 
the resolution $\delta_{\hat s}$ enters:
\begin{equation}
\label{Eq:R-RS}
R^{\rm RS}
=\frac{9}{2048}\, \frac{(\kappa m_G)^2}{\alpha^2}\,
\frac{1}{\delta_{\hat s} \gamma_G}\, 
\frac{I_{gg}(m_G^2)}{I_{q\bar q}(m_G^2)}\,
\frac{\tilde\sigma^{(G)}_{gg\to\mu^+\mu^-\gamma}}
{\tilde\sigma^{(\gamma)}_{q\bar q\to\mu^+\mu^-\gamma}}.
\end{equation}

In order to estimate the total number of $\mu^+\mu^-\gamma$ events 
produced, we consider the sum of QED and graviton-induced cross
sections in Figs.~\ref{Fig:add} and \ref{Fig:rs}.
Here, we have chosen a $y$-cut of 0.01, and consider a minimum 
$x_3$ of 0.1\footnote{These cuts have not been optimized.}.

%%%%%%%%%%%%%%%%%%%%%%%%%%%%%%%%%%%%%%%%%%%%%%%%%%%%%%%%%%%%%%%%%%%%%%
\begin{figure}[htb]
\refstepcounter{figure}
\label{Fig:add}
\addtocounter{figure}{-1}
\begin{center}
\setlength{\unitlength}{1cm}
\begin{picture}(14,7.3)
\put(-1.2,0.0)
{\mbox{\epsfysize=8.0cm\epsffile{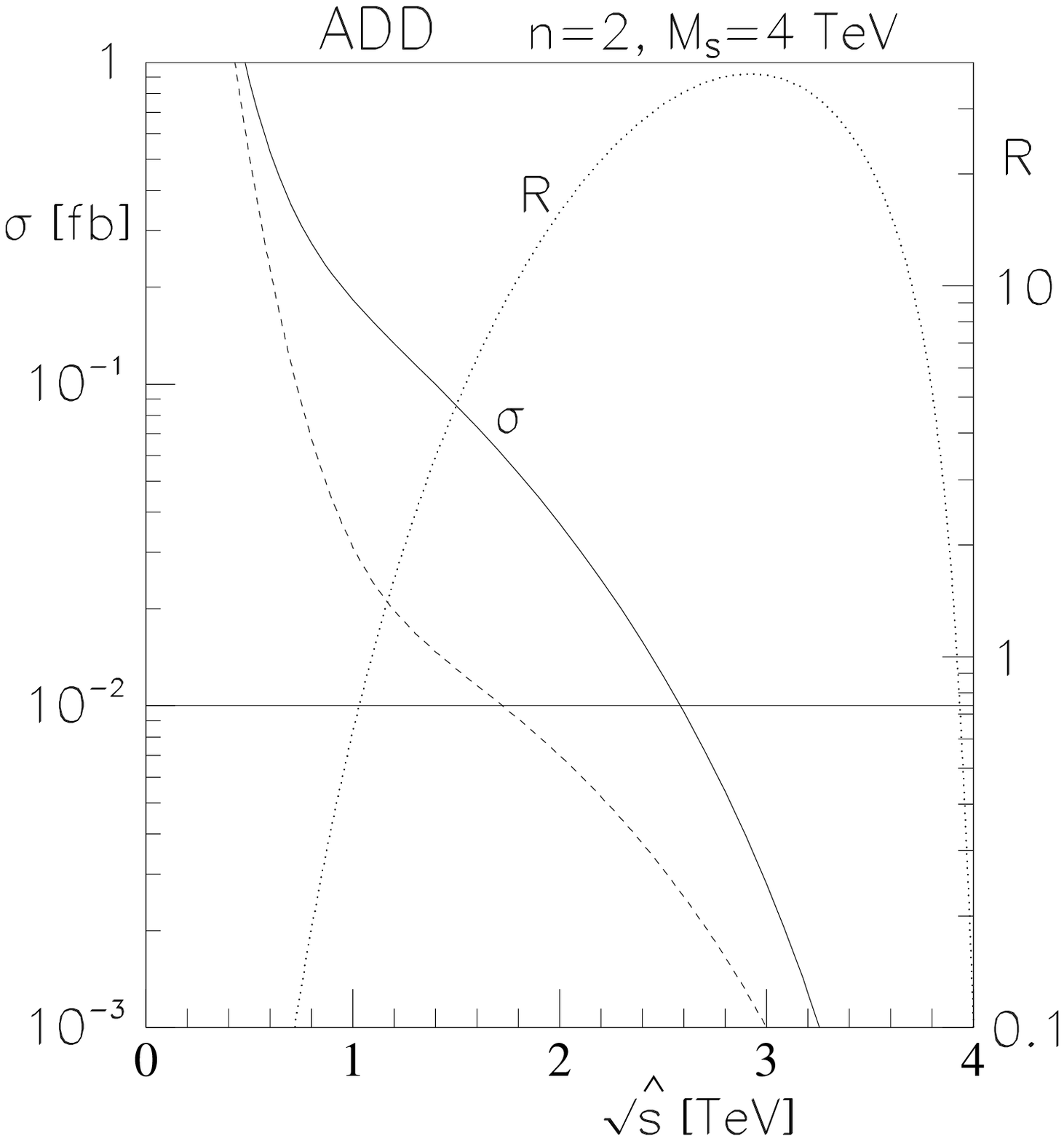}}
 \mbox{\epsfysize=8.0cm\epsffile{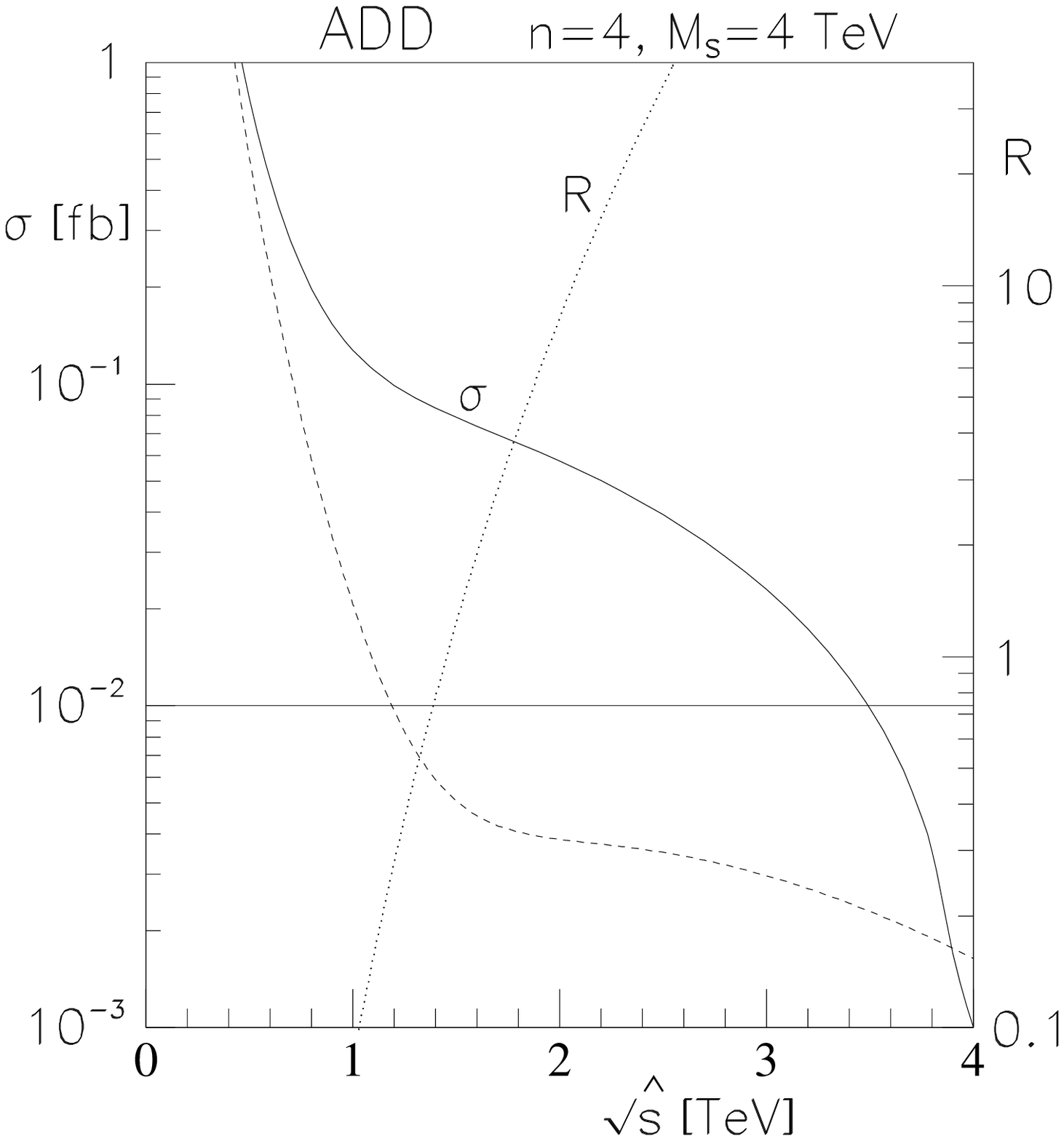}}}
\end{picture}
%\vspace*{-8mm}
\caption{ADD case, $n=2$ and $n=4$, $M_S=4$~TeV.
Differential cross section,
$100\times d\sigma/d\sqrt{\hat s}$ (dashed) in [fb/GeV],
integrated cross section, $\sigma(\sqrt{\hat s})$ (solid), 
Eq.~(\ref{Eq:sig-add-int}),
and signal to QED ratio, $R$ (dotted).}
\end{center}
\end{figure}
%%%%%%%%%%%%%%%%%%%%%%%%%%%%%%%%%%%%%%%%%%%%%%%%%%%%%%%%%%%%%%%%%%%%%%

For the ADD scenario (Fig.~\ref{Fig:add}), we plot 
$d\sigma/d\sqrt{\hat s}$ according to Eqs.~(\ref{Eq:dsigma-gg-add}),
(\ref{Eq:dsigma-DY-pp}), (\ref{Eq:sig-G-ints})
and (\ref{Eq:sig-ga-ints}) (dashed). These curves are scaled
by a factor of 100, and thus give the cross section per 100~GeV bin.
Also shown, as solid curves, are the integrated cross sections:
\begin{equation}
\label{Eq:sig-add-int}
\sigma(\sqrt{\hat s})
=\int_{\sqrt{\hat s}}^{M_S} d\sqrt{\hat s'}\, 
\frac{d\sigma}{d\sqrt{\hat s'}}.
\end{equation}
The line at $10^{-2}~\text{fb}$ corresponds to 1 event per year,
at the integrated LHC luminosity of $100~\text{fb}^{-1}$ 
(one nominal LHC year at $L=10^{34}~\text{cm}^{-2}\,\text{s}^{-1}$).
Finally, the dotted curve labelled $R$ shows the ``figure-of-merit'',
Eq.~(\ref{Eq:R-ADD}).
We consider two values for the number of extra dimensions, $n=2$ and 4, and
take the string scale, $M_S$, to be $4~\text{TeV}$.
It is seen that there will be a significant number of such
Bremsstrahlung events up to a few TeV,
and that for the larger values of $\sqrt{\hat s}$, these tend
to be gravity-dominated.

%%%%%%%%%%%%%%%%%%%%%%%%%%%%%%%%%%%%%%%%%%%%%%%%%%%%%%%%%%%%%%%%%%%%%%
\begin{figure}[htb]
\refstepcounter{figure}
\label{Fig:rs}
\addtocounter{figure}{-1}
\begin{center}
\setlength{\unitlength}{1cm}
\begin{picture}(14,7.3)
\put(-1.2,0.0)
{\mbox{\epsfysize=8.0cm\epsffile{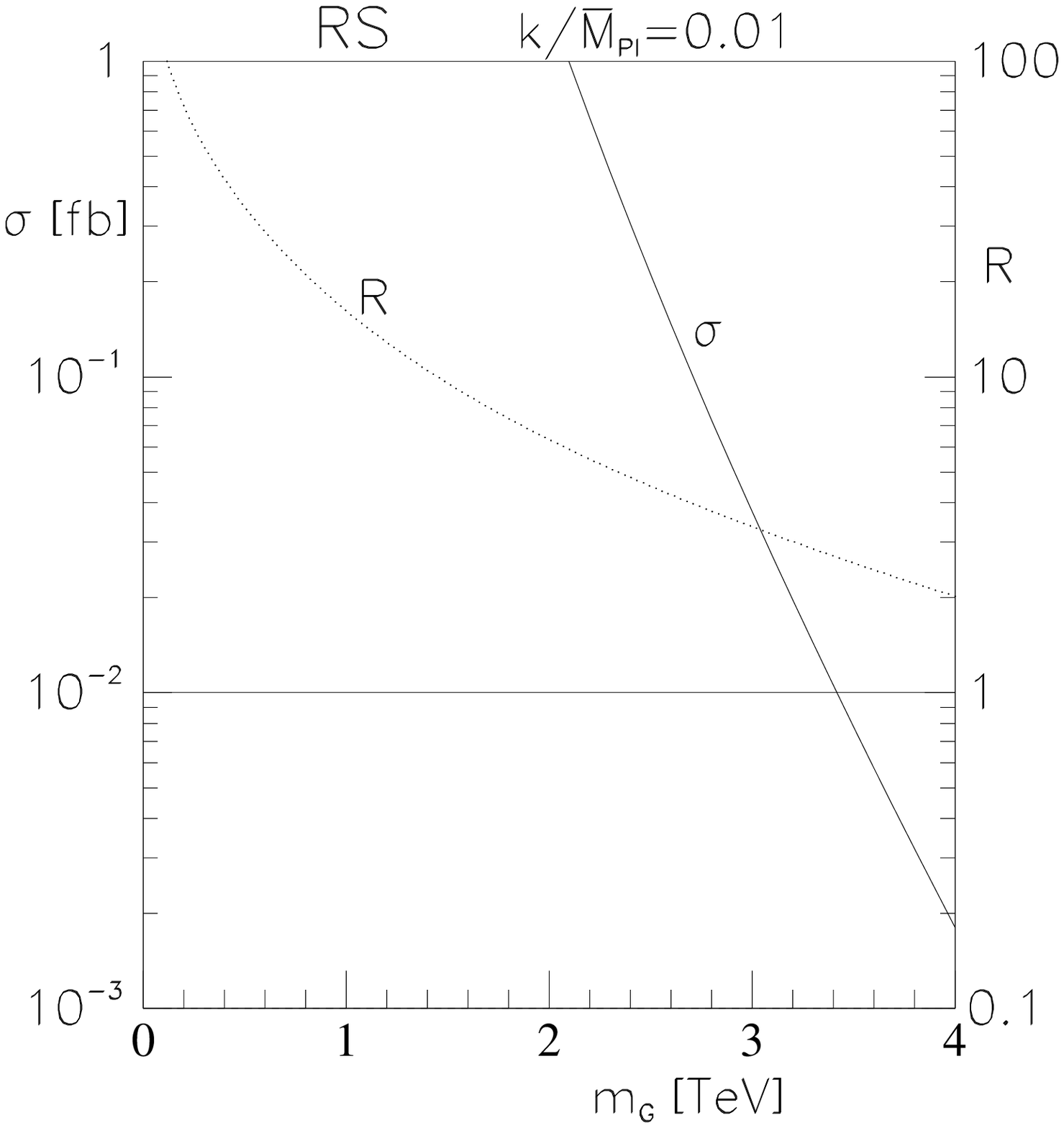}}
 \mbox{\epsfysize=8.0cm\epsffile{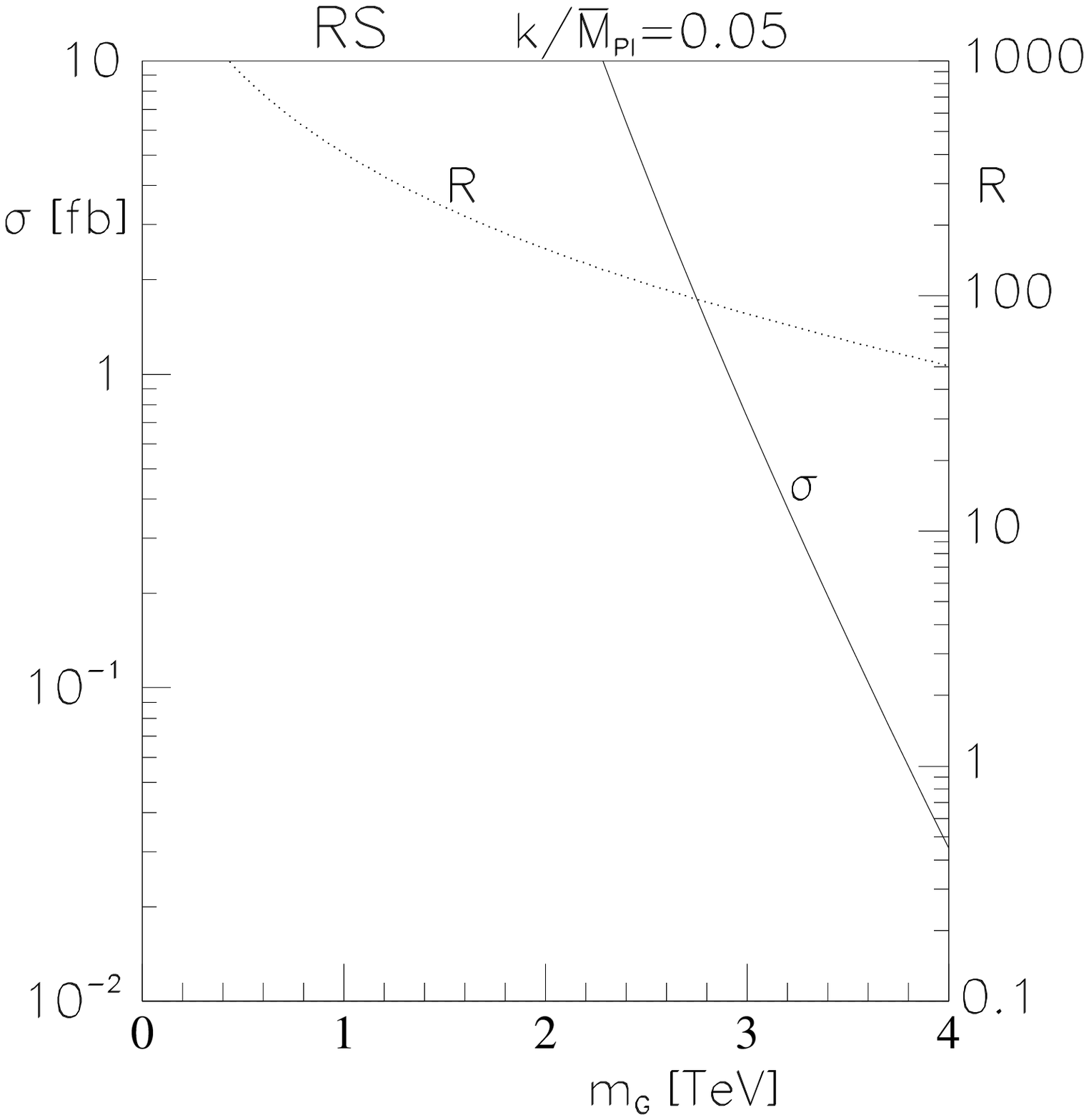}}}
\end{picture}
%\vspace*{-8mm}
\caption{RS case, $k/\overline M_{\rm Pl}=0.01$ and 0.05.
Cross sections and signal to QED ratio.
The graviton-exchange cross section is evaluated in the narrow-width
approximation, with the resolution $\delta_{\hat s}=0.10$.}
\end{center}
\end{figure}
%%%%%%%%%%%%%%%%%%%%%%%%%%%%%%%%%%%%%%%%%%%%%%%%%%%%%%%%%%%%%%%%%%%%%%

In the RS case, we consider $k/\overline M_{\rm Pl}=0.01$ and 0.05,
and a resolution roughly approximated as $\delta_{\hat s}=10\%$.
We display the corresponding signal plus QED background
cross sections, given by Eqs.~(\ref{Eq:sig-rs-int}) and 
(\ref{Eq:sig-G-ints}) (divided by $\Delta \hat s$), together with 
(\ref{Eq:sigma-DY-pp}) and (\ref{Eq:sig-ga-ints}) (solid).
Also shown, is the ratio of gravity-induced
to QED cross section, as given by the ``figure of merit'', 
Eq.~(\ref{Eq:R-RS}).

In summary,
we have discussed Bremsstrahlung induced by the exchange of massive 
gravitons at hadron colliders, in particular at the LHC. 
Both the ADD and the RS scenarios have been considered.
We found that three-body final states can be a valuable
supplement to the two-body final states, for the purpose
of detecting massive gravitons related to extra dimensions. 
These configurations, of a hard photon associated with a muon pair
in the opposite direction, should provide a striking signal
at the LHC.

We have here focused on gluon fusion.
There is also graviton exchange induced by quark-antiquark annihilation.
These contributions are of importance at larger invariant masses,
and will be discussed elsewhere \cite{DOO-2}. It could also be interesting to
consider similar processes where the final-state fermions are massive.

%%%%%%%%%%%%%%%%%%%%%%%%%%%%%%%%%%%%%%%%%%%%%%%%%%%%%%%%%%%%%%%%%%%%%%%%
\bigskip

{\bf Acknowledgements.}
It is a pleasure to thank Paolo di Vecchia for very useful discussions.
This research has been supported by the Research Council of Norway
and by NORDITA.

\goodbreak

\end{document}